\documentstyle[11pt,psapconf,epsfig]{article}

\begin{document}

\title{Damped Ly$\alpha$ Systems in Semi-Analytic Models: Sensitivity
to dynamics, disk properties, and cosmology.}

\author{Ariyeh Maller}
\affil{Physics Department, University of California,
    Santa Cruz, CA 90564}

\author{Rachel S. Somerville}
\affil{Racah Institute of Physics, The Hebrew University, Jerusalem, 91904, 
Israel}

\author{Jason X. Prochaska}
\affil{Observatories of the Carnegie Institution of Washington, Pasadena, CA 
91101}

\author{Joel R. Primack}
\affil{Physics Department, University of California,
    Santa Cruz, CA 90564}

\keywords{quasars: absorption lines,galaxies: formation,
galaxies: kinematics and dynamics}

\section{Introduction}
Previously (Maller et~al. 1999, hereafter MSPP) we have shown that it is
possible to account for the kinematic properties of damped Lyman alpha
systems (DLAS) as measured by Prochaska \& Wolfe 
(1997, 1998, hereafter PW97 and
PW98) in the context of semi-analytic models (SAMs)
(Somerville \& Primack 1999, hereafter SP99).  
In these models, hierarchical structure formation is
approximated by constructing a merger tree for each dark matter halo.
A natural consequence is that every virialized halo may contain not
only a central galaxy, but also a number of satellite galaxies as
determined by its merging history.  Thus the kinematics of the DLAS
arise from the combined effects of the internal rotation of gas disks
and the motions between gas disks within a common halo.  Here we
investigate the sensitivity of this model to some of the assumptions
made in MSPP, including the modeling of satellite dynamics, the scale
height of the gas, and the cosmology.

\section{Satellite Dynamics}
In the SAMs, a merger tree represents a halo's growth through the
mergers of smaller halos (see {Somerville} 1997). 
When halos merge, the
central galaxy of the largest progenitor halo becomes the central
galaxy of the new halo, and all other galaxies become
satellites. These satellites then fall in towards the central galaxy
due to dynamical friction, and eventually merge with it (see SP99). 
Because the treatment of the dynamics of the
satellites is necessarily simplified in the usual semi-analytic
spirit, and since the kinematics of the DLAS arise from both the
rotation of disks and the motions of satellite galaxies in the common
halo, it is important to test whether our results are sensitive to the
details of our modeling.

\begin{figure}[t]
\vskip .1pc
\begin{minipage}[b]{.46\linewidth}
\centering
\centerline{\epsfig{file=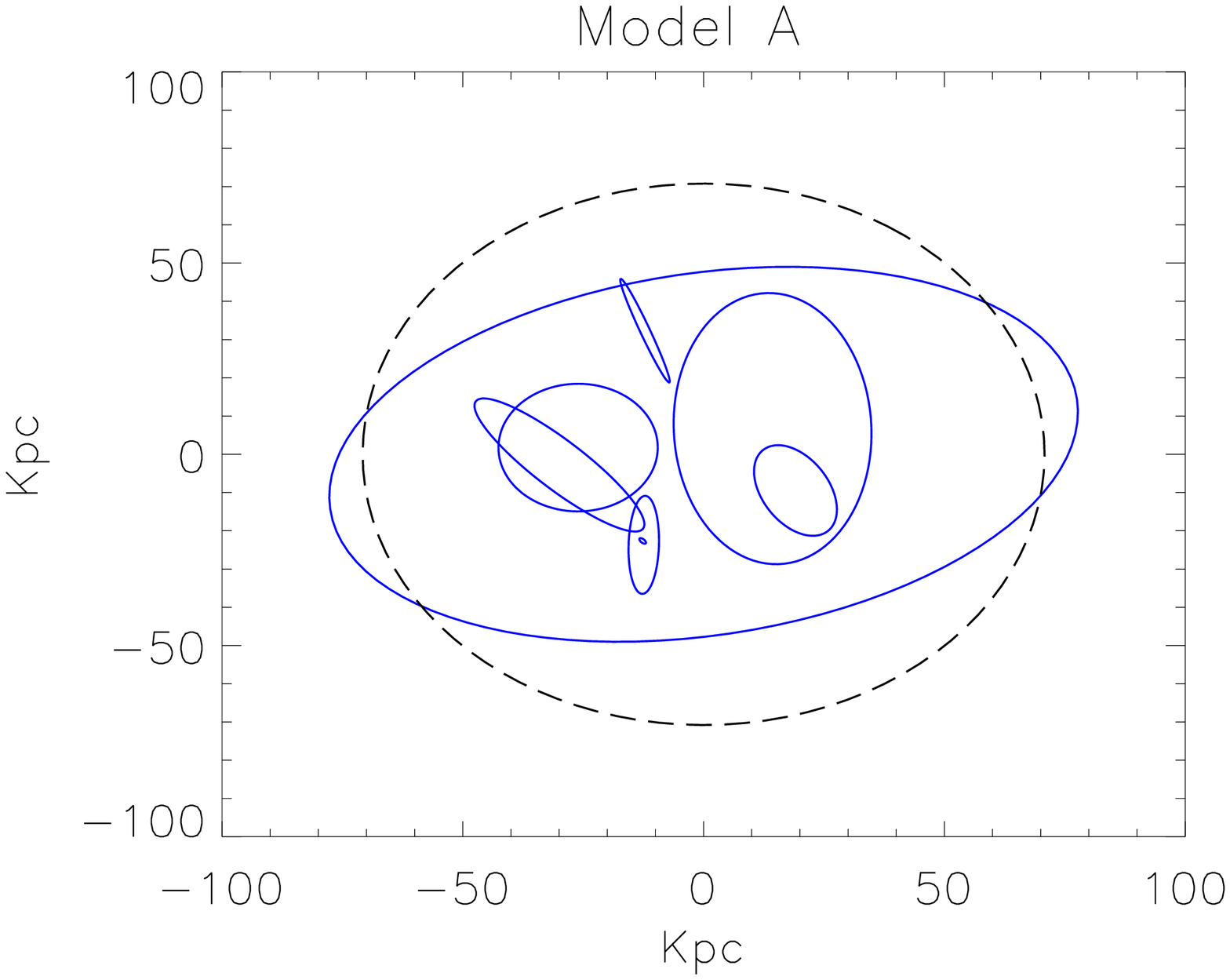,height=.8\linewidth,width=.9\linewidth}}
\end{minipage}\hfill
\begin{minipage}[b]{.46\linewidth}
\centering
\centerline{\epsfig{file=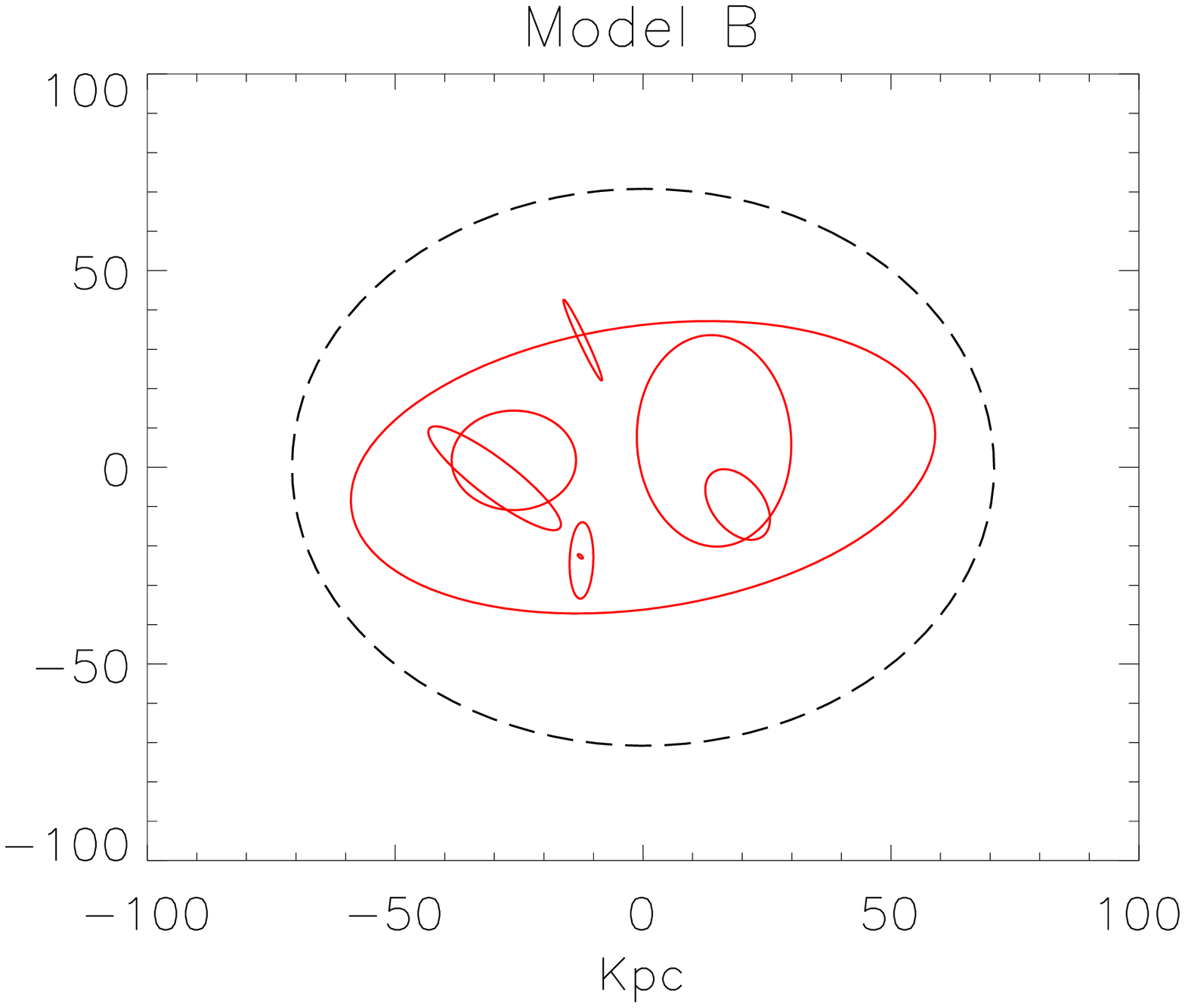,height=.8\linewidth,width=.9\linewidth}}
\end{minipage}\hfill
\vskip .1pc
\caption{A halo of circular velocity 156 km s$^{-1}$ with its 
satellite galaxies.  The ellipses mark where the cold gas is truncated, 
in Model A (left) with thin disks and Model B (right) with thicker disks.  
The dashed line is the virial radius of the parent halo.
}\label{pic}
\end{figure}

In the models presented in MSPP, all satellites were assumed to start
falling in from the virial radius of the newly formed halo, and all
satellites were assumed to be on circular orbits. These assumptions
will tend to maximize the dynamical friction timescale of the
satellites, and correspond to the assumptions made in some earlier
versions of the SAMs (e.g. Somerville 1997). In the models
presented here, we have modified these assumptions in two ways. We
initially place the satellites at one half of the virial radius, and
assign the orbits from a random distribution as observed in N-body
simulations by {Navarro}, {Frenk}, \& {White} (1995). 
They found that the ``circularity''
parameter $\epsilon$ has a flat distribution, where $\epsilon \equiv
J/J_c$ is defined as the ratio of the angular momentum of the
satellite to that of a circular orbit with the same energy. The
dynamical friction time scale is then scaled by a factor
$\epsilon^{0.78}$ ({Lacey} \& {Cole} 1993). 
This causes some of the satellites to
fall in faster than in our previous modeling. In addition, we have
changed from the SCDM cosmology used in MSPP to a more fashionable
$\Omega_0=0.4$, $\Omega_{\Lambda}=0.6$ cosmology. These ingredients
are compatible with the models that were shown to produce good
agreement with both local galaxy observations (SP99) 
and the high redshift Lyman-break galaxies 
({Somerville}, {Primack}, \&  {Faber} 1999).

We find that the combined effect of these changes has a negligible
effect on our results. Although we do see fewer satellites within a
halo of a given mass, the number of satellites in the inner part of
the halo is similar. These are the satellites most likely to give rise
to multiple hits, which as we argued in MSPP is the crucial factor in
matching the observed kinematics of the DLAS. As in MSPP, we still
find that the gaseous disks must have very large radial extents in
order to match the observations.

\section{Disk Thickness}
Another important feature of the models is the assumed distribution of
the gas within the disks. In MSPP, we investigated several radial
profiles and obtained the best results with an assumed gas
distribution of the form
\begin{equation}
n(R,z)= {{N_t} \over {2 h_z}}{R_t \over R} \exp{ {{-|z|}\over {h_z}}}  
\hskip .5in   (R < R_t)
\end{equation}
where the truncation density $N_t$ is an adjustable parameter.  The
vertical scale height of the gas disks is another uncertainty. In MSPP
we assumed it to be one tenth of the stellar disk scale radius,
i.e. $h_z= 0.1 R_*$ where $R_*$ is the stellar disk scale length as
given by the SAMs (see SP99). Because the gas disks in the
successful models tend to have a large radial extent, much larger then
the stellar disk, this leads to very thin gaseous disks. When the
scale height is increased to half the stellar scale radius ($h_z=0.5
R_*$), we are able to use more physically plausible values for the
truncation density $N_t$ (see Table~\ref{modelab}), and the gas disks
are now typically contained within the truncation radius of the dark
matter sub-halos surrounding the satellites. This model therefore
seems to be more physical, yet still produces good agreement with all
four of the diagnostic statistics of PW97 (see
Table~\ref{modelab}). Note that PW98 have shown that the effect of
including a warp in the gas disk is the same as increasing its
thickness, so our exploration of thicker disks can also be thought of
as including warps in the disk.

\begin{table} [t]
\center
\begin{tabular}{lclcccc}
\tableline
Model & vertical scale height & normalization & 
$\Delta V$ & f$_{mm}$ & f$_{edg}$& f$_{tpk}$\\
\tableline
\tableline
A  & .1$R_{*}$ &  
$\log N_t=19.3$ & 0.35 & 0.10 & 0.56 & 0.26 \\
B  & .5$R_{*}$  &
$\log N_t=19.6$  &  0.41 & 0.24 & 0.54 & 0.71 \\
\tableline
\end{tabular}
\caption{The properties of our two models and the KS probabilities for
the four statistics of PW97. } 
\label{modelab}
\end{table}

\begin{figure}
\vskip .5pc
\begin{minipage}[b]{.46\linewidth}
\centering
\centerline{\epsfig{file=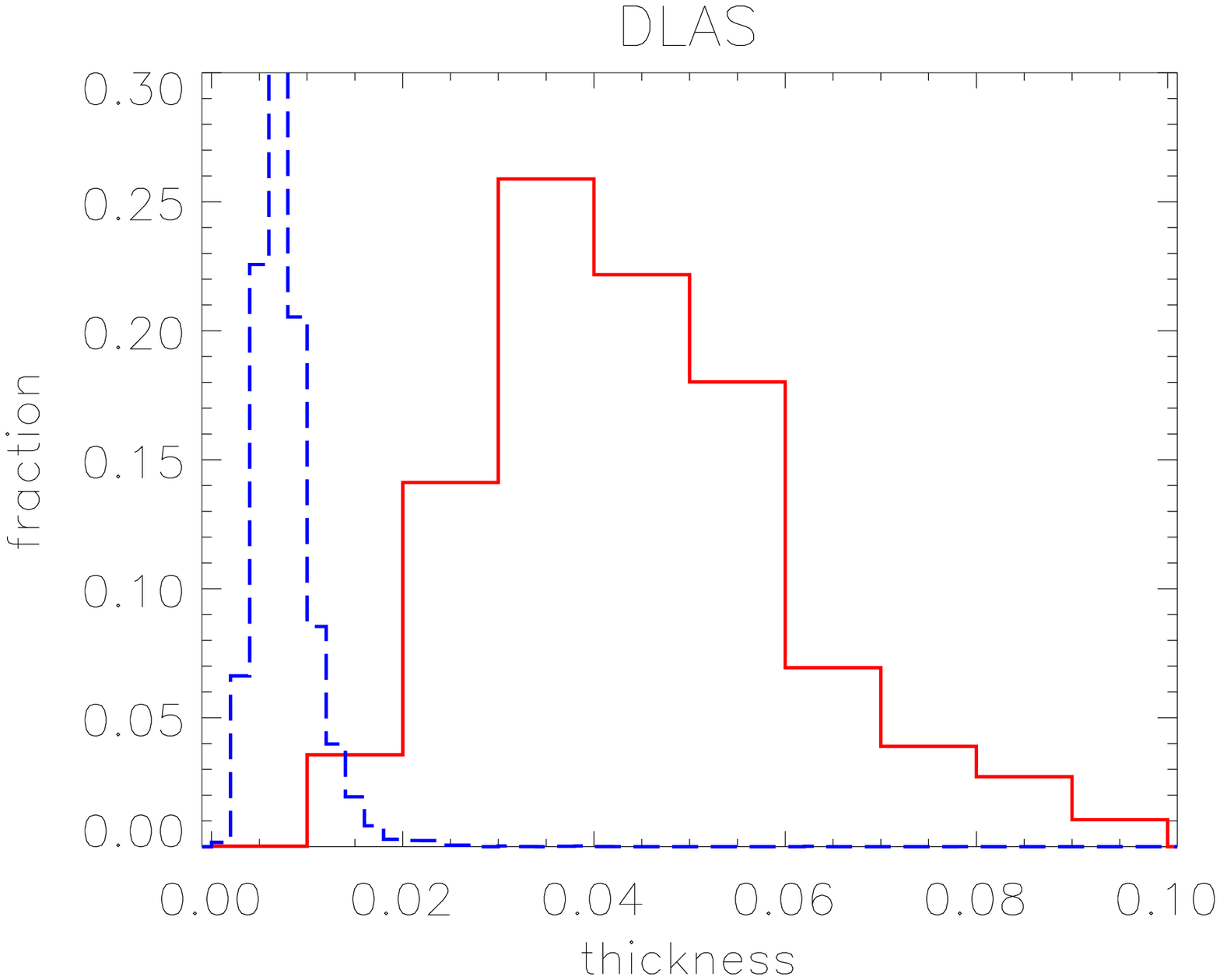,width=\linewidth}}
\end{minipage}\hfill
\begin{minipage}[b]{.46\linewidth}
\centering
\centerline{\epsfig{file=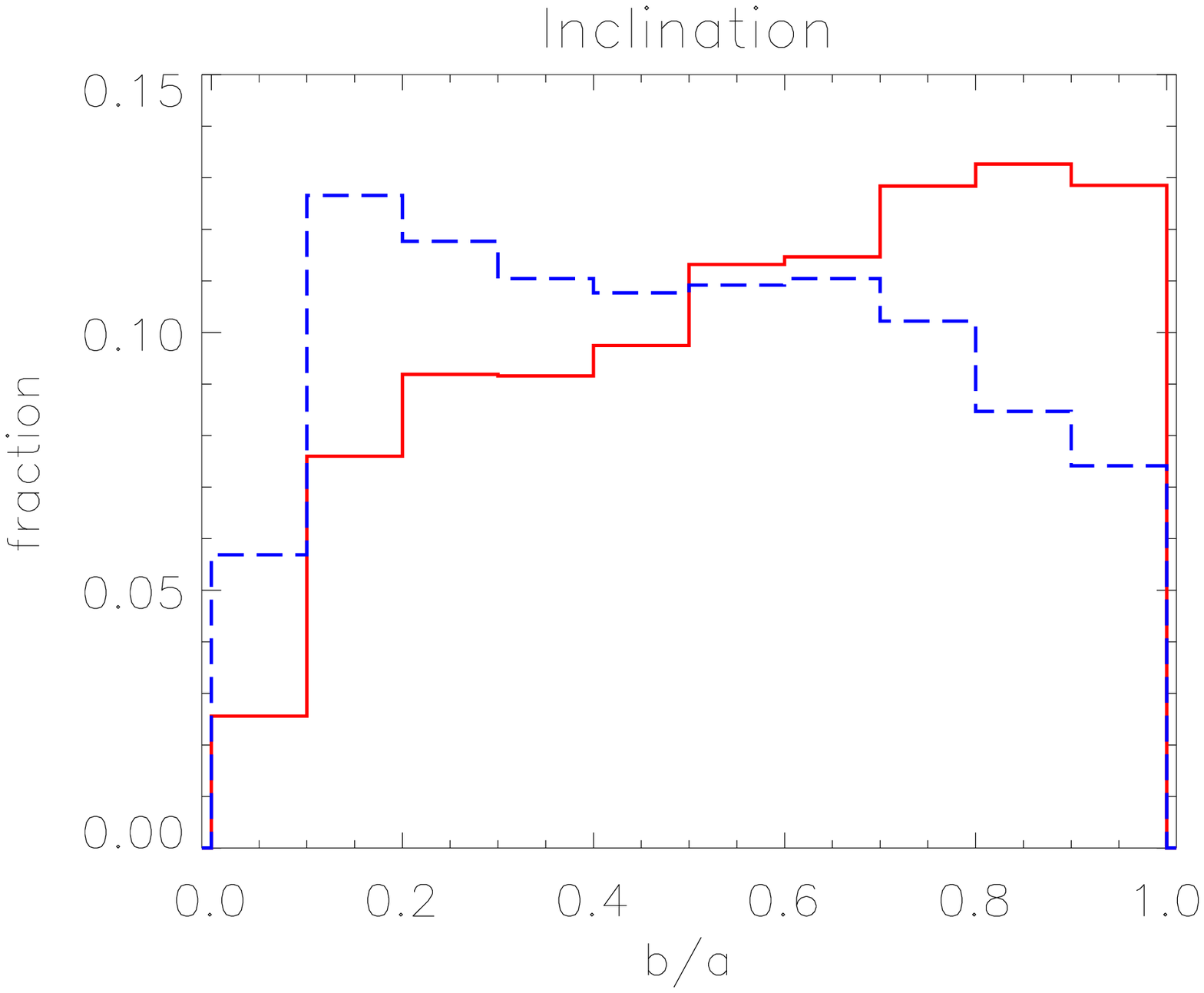,width=\linewidth}}
\end{minipage}\hfill
\vskip .5pc
\caption{The distribution of thicknesses (left panel) and inclinations 
(right panel) of the disks that produce DLAS from our two models; 
Model A (dashed), Model B (solid).
We characterize the thickness as $3h_z/R_t$. 
The DLAS in Model A are produced by more inclined disks then in Model B. 
}\label{figthick}
\end{figure}

Figure \ref{figthick} shows the distribution of thicknesses and
inclinations for the disks that contribute to DLAS in the two models.
We find that the disks that produce damped systems in model B are more
likely to have a face-on geometry.  With thinner disks, more of the
cross section comes from an inclined geometry as $n \sim h_x^{-1}$.
The total cross section in the two models is roughly the same; in
model A, more of it comes from extended highly inclined disks, while
in model B, denser, face-on disks are more important.

Thus the need for gas disks with large radial extent can be reduced by
using thicker disks, though not by a large amount.  Increasing the
thickness by a factor of five only reduces the radial truncation value
by $30\%$. The gaseous disks in model B are still quite large compared
to the stellar component.
  
\section{Conclusions}
In MSPP we argued that the observed kinematics of DLAS can be
reproduced in hierarchical models if a significant fraction of the
lines of sight pass through multiple disks orbiting within a common
dark matter halo. However, we found that in order to obtain a high
enough cross section for multiple hits, we had to assume very large
radial extents for the gaseous disks in our models.  We have tested
the robustness of these conclusions by modifying several of the
uncertain ingredients of our models. We find that modifying the
dynamical friction timescale of the satellites by assuming a different
intial radius or orbit has a small effect on our results. Similarly,
our results do not seem to be sensitive to the assumed
cosmology. Increasing the vertical scale height of the disks has a
larger effect and leads to models that are more physically plausible
and still produce good agreement with the diagnostic statistics of
PW. Our results suggest that in order to reconcile the observed
kinematics of DLAS with hierarchical theories of structure formation,
gaseous disks at high redshift must be large in radial extent and
thickened or warped. The physical cause of these properties remains
obscure, however we speculate that tidal encounters or outflows could
be responsible.

\end{document}